\documentstyle[11pt]{article}
\begin{document}
\begin{flushright}SJSU/TP-98-18\\ September 1998\end{flushright}
\vspace{1.7in}
\begin{center}\Large{\bf Ambiguities of arrival-time distributions \\
   in quantum theory\\} 
\vspace{1cm}
\normalsize\ J. Finkelstein\footnote[1]{
        Participating Guest, Lawrence Berkeley National Laboratory\\
        \hspace*{\parindent}\hspace*{1em}
        e-mail: JLFinkelstein@lbl.gov}\\
        Department of Physics\\
        San Jos\'{e} State University\\San Jos\'{e}, CA 95192, U.S.A
\end{center}
\begin{abstract}
 We consider the definition that might be given to the time at which 
 a particle arrives at a given place, both in standard quantum theory and 
 also in Bohmian mechanics. We discuss an ambiguity that arises in the 
 standard theory in three, but not in one, spatial dimension.
\end{abstract}
\newpage
\section{Introduction}
In classical mechanics, a particle can be said to follow a definite 
trajectory, and so it is clear what is meant by the time at which a 
particle arrives in a given place.  If one considers an ensemble of 
particles, it is then easy to say what is meant by a distribution of arrival
times.  In standard quantum theory, on the other hand, particles are not 
said to follow trajectories, and so the meaning of arrival time in quantum 
theory has been rather controversial.  Some elements of this controversy
include statements such as ``the time-of-arrival cannot be precisely defined
and measured in quantum mechanics'' (quoted from \cite{aha}; see also
\cite{Allcock, Mielnik, HZ}); statements that time-of-arrival {\em must} be
definable, for example ``Since the distribution of arrival
times at a given spatial point is, in principle, a measurable quantity
that can be determined via a time-of-flight experiment, it is reasonable
to ask for an apparatus-independent theoretical prediction'' 
 \cite{Del1}; specific proposals for defining a time-of-arrival
distribution (for example, in \cite{ MBM, GRT, DM}), and suggestions
(for example, in \cite{GRT} and \cite{Del2})
that experiments could determine which if any of these proposals is correct.

In the causal theory of Bohm \cite{Bohm, BH}, particles {\em do} follow 
definite trajectories, and so the definition of arrival-time distributions 
is again unambiguous. Leavens, most recently in \cite{Leavens}, has 
studied the 
arrival-time distribution of a free particle in Bohmian theory, and found 
results which differ from the proposal made in \cite{GRT}.  Deotto and 
Ghirardi
\cite{DGh}, and also Holland \cite{Holland} have proposed what I shall
call Bohm-like theories: theories in which particles follow trajectories
which differ from the trajectories of Bohmian theory, but which 
nevertheless reproduce all of the observational results of standard
quantum theory, in the same way that Bohmian theory does. In this paper I will
study a simple example of a Bohm-like theory, and demonstrate that in 
certain cases this theory will produce arrival-time distributions
which are different that those produced by (standard) Bohmian theory. 
I will also use
this same demonstration to discuss, quite apart from any consideration
of Bohmian theory, an ambiguity in the meaning of arrival-time distributions
for three-dimensional problems within standard quantum theory.

I begin by using terminology which is appropriate is classical mechanics.
Consider a particle with position coordinates $(x,y,z)$, which at the 
initial time $t=0$ has $x<0$. Let $S_+$ denote the region of space in which
$x\geq 0$, $S_-$ the region in which $x<0$, and $S_0$ the $x=0$ plane. Let 
$T$ be the first time after $t=0$ at which the particle arrives at $S_0$
(equivalently, crosses from $S_-$ to $S_+$); by definition, $T>0$. I will 
mostly be concerned with the integrated arrival-time distribution,
which I denote as $P(t)$; that is, $P(t)$ is the probability that $T\leq t$. 

The question to be discussed is whether this distribution $P$ can be
precisely and unambiguously defined in quantum theory.  If it can be, 
it could be expected to satisfy, at least, the following properties:

\begin{description}
\item[i)] $P(t)$ is monotonically increasing: $P(t) \geq P(t')$ for
  $t\geq t'\geq 0$.
\item[ii)] Define $A(t)=dP(t)/dt$; from i), we have $A(t)\geq 0$.
  Then $A(t)dt$ represents the probability that $T=t$. 
\item[iii)] Let $\overline{T}$ be the average value of $T$ 
  (averaged over those cases in which the particle does eventually 
  arrive at $S_0$). From ii), this is given by
  \begin{equation}
  \overline{T}=\frac{\int_{0}^{\infty}t\,A(t)\,dt}{\int_{0}^{\infty}A(t)\,dt}
  \end{equation}
  This can be re-written, with 
  $P_{\infty }:=\lim_{t\rightarrow \infty}P(t )$, as
  \begin{equation}
    \overline{T}=\frac{\int_{0}^{\infty}dt\,[P_{\infty }-P(t)]}{P_{\infty }}
  \end{equation}
\item[iv)] Now let $Q(t)$ be the probability that the particle would be
  found, at time $t$, in $S_+$; that is,
  \begin{equation}
     Q(t)=\int \int \int|\Psi(x,y,z,t)|^{2}\,\theta (x)\,dx\,dy\,dz.
  \label{ez}\end{equation}
  The initial condition we are assuming means that $Q(t=0)=0$. We expect that
  \begin{equation}P(t)\geq Q(t).\label{i1}\end{equation}
  If the particle could surely be found in $S_+$ at time $t$
  if it had arrived there at time $t'$ (with $0\leq t'\leq t$), then the
  relation (\ref{i1}) could be replaced by an equality.
\item[v)] From i) and iv), we have
  \begin{equation}P(t)\geq \max_{t\geq t'\geq 0}Q(t').\end{equation}
\end{description}

In the next section I will present an example in which a Bohm-like theory
produces a different result for $P(t)$ than does the usual Bohmian theory.
The implications of this example are discussed in the final section.

\section{Example}
Consider a free ($V=0$) particle which, at the initial time $t=0$, is 
described by
the following minimum-uncertainty wave packet centered at the point
$x=-x_1$, $y=0$, $z=0$:
\begin{eqnarray}
\Psi(x,y,z,t=0)&=&\left[\frac{1}{\pi^{3}a^{2}b^{2}c^{2}}\right] ^{\frac{1}{4}} 
   \exp[ikx]\exp\left[-\frac{(x+x_{1})^{2}}{2a^{2}}\right]\nonumber\\
   & &\mbox{}\times \exp\left[-\frac{y^{2}}{2b^{2}}\right]
   \exp\left[-\frac{z^{2}}{2c^{2}}\right],\label{e1}\end{eqnarray}
where $a$, $b$, $c$, and $k$ are positive constants, and where I
have set both the mass of the particle and the value of $\hbar$ to one.
We want the particle to start out with $x< 0$; this corresponds to
$x_{1}> 0$. Strictly speaking, this wave packet does not
satisfy the condition $Q(t=0)=0$, because its tail extends to positive
values of $x$. However, $Q(0)$ can be made arbitrarily
small by taking $(x_{1}/a)$ large (see Eq.~\ref{ew} below).

Define $\alpha =(a^{2}+it)^{\frac{1}{2}}$;
$\beta =(b^{2}+it)^{\frac{1}{2}}$; $\gamma =(c^{2}+it)^{\frac{1}{2}}$.
Then 
\begin{eqnarray}
  \Psi(x,y,z,t)&=&\left[\frac{a^{2}b^{2}c^{2}}{\pi ^{3}}\right]^{\frac{1}{4}}
   \frac{\exp[i(kx-k^{2}t/2)]}{\alpha \beta \gamma}
   \exp\left[-\frac{(x+x_{1}-kt)^{2}}{2\alpha ^{2}}\right]\nonumber\\
   & &\mbox{}\times \exp\left[-\frac{y^{2}}{2\beta ^{2}}\right]
   \exp\left[-\frac{z^{2}}{2\gamma ^{2}}\right].\label{e2}\end{eqnarray}
At time $t$, the center of the wave packet is at $x=-x_{1}+kt$ (with
$x_{1}>0$ and $k>0$), $y=z=0$.
It is straightforward to calculate
\begin{equation} 
  Q(t)=\mbox{\small $\frac{1}{2}$}{\rm erfc}(\eta), \label{e3}
  \end{equation}
where $\eta:=a(x_{1}-kt)/|\alpha |^2$, and erfc is the complementary
error function: erfc$(\eta )=2\pi ^{-\frac{1}{2}}\int_{\eta}^{\infty}
\exp (-\xi ^{2})\,d\xi $.  It follows from Eq.~\ref{e3} that
\begin{equation}
  Q(0)=\mbox{\small $\frac{1}{2}$}{\rm erfc}(x_{1}/a),
  \label{ew}\end{equation}
and that
\begin{equation}
  \lim_{t\rightarrow \infty}Q(t)=\mbox{\small $\frac{1}{2}$}
  {\rm erfc}(-ak).\label{ey}\end{equation}
We see from (\ref{ey}) that $Q$ does not approach one even as $t\rightarrow
\infty$; this is because the wave packet spreads out as its center moves
toward large $x$, so that a finite fraction of the tail of the packet
remains in $S_-$.

In Bohmian theory \cite{Bohm, BH} the particle is considered to have a 
definite position, which will be denoted as ${\bf r}=(X,Y,Z)$. Let
${\bf v}_b$ denote the Bohmian velocity of the particle (that is, 
${\bf v}_{b}=d{\bf r}/dt)$; then if we write $\Psi =R\exp(iS)$, in
Bohmian theory ${\bf v}_b$ is given by
\begin{equation}
  {\bf v}_b = {\bf \nabla} S \label{e4}\end{equation}
It then follows that
\begin{equation}
  {\bf \nabla}\cdot (|\Psi |^{2}{\bf v}_{b})=-\frac{\partial |\Psi |^{2}}
   {\partial t} \label{e5}\end{equation}
In fact, the product $(|\Psi |^{2}{\bf v}_{b})$ is just the usual quantum
probability current, and Eq.~\ref{e5} is just the 
equation of conservation of probability in standard quantum theory.
One associates with a given wave function $\Psi$ an ensemble of particles,
whose distribution agrees with the quantum probability density
$|\Psi|^2$; Eq.~\ref{e5} assures that this agreement, if it exists at
the initial time, persists for all time, and this in turn means that
Bohmian theory will reproduce all of the experimental predictions of
standard quantum theory \cite{ex}.  Thus Bohmian theory is not in 
conflict with, but rather is a completion of, standard quantum theory.

Since each particle in the Bohmian ensemble follows a definite trajectory,
the interpretation of arrival-time distributions is unambiguous.
The quantity $P(t)$ defined above is simply the fraction of particles in
the ensemble which have $X\geq 0$ for any time $t'$ with $0\leq t' \leq t$,
and of course $Q(t)$ is the fraction which have $X\geq 0$ at time $t$.
For the wave function in Eq.~\ref{e2}, the components of the Bohmian
velocity turn out to be
\begin{equation}
 v_{bx}(X,Y,Z,t)=[k+(X+x_{1})t]/|\alpha |^{4},\label{e6}\end{equation}
\begin{equation}
 v_{by}(X,Y,Z,t)=Yt/|\beta |^{4},\label{e7}\end{equation}
\begin{equation}
 v_{bz}(X,Y,Z,t)=Zt/|\gamma |^{4},\label{e8}\end{equation}
Since $x_{1}>0$, we see from Eq.~\ref{e6} that, for all $t\geq 0$,
\begin{equation}
 v_{bx}(X=0,Y,Z,t)>0\label{e9}\end{equation}
This means that if the particle does enter $S_+$, it can 
never leave.  Since that was the condition which gives equality in
the relation (\ref{i1}), we see that in this example, Bohmian theory gives
\begin{equation}
  P(t)=Q(t), \label{e10}\end{equation}
where $Q(t)$ is given by Eq.~\ref{e3}.

Because of the factorized form of $\Psi$ in this example, 
the x-component of the Bohmian motion is the same as in the one-dimensional
example of a minimum-uncertainty packet studied in \cite{Leavens}. In fact
we can, without having to solve for the Bohmian trajectories in detail,
recover one of the main results of \cite{Leavens}, namely that a finite
fraction of the Bohmian ensemble never makes it to the region $S_+$.
That fraction is just $(1-P_{\infty})$; by
Eq.~\ref{e10} this equals  $(1-\lim_{t\rightarrow \infty}Q(t))$,
which we saw in Eq.~\ref{ey} is not zero.
  
It is Eq.~\ref{e5} which insures, for Bohmian theory, 
that an ensemble of particles with
initial distribution given by $|\Psi |^2$ reproduces the
experimental predictions of standard quantum theory.  One can formulate
an alternative theory, which I will refer to as a Bohm-like theory,
in which a particle again has a definite position, but in which the
velocity (call it ${\bf v}_{bl})$ may differ from the Bohmian velocity
${\bf v}_b$ (given in Eq.~\ref{e4}).  Let $\delta {\bf v}$ denote
the difference between ${\bf v}_{bl}$ and ${\bf v}_b$:
\begin{equation}
   {\bf v}_{bl}= {\bf v}_{b}+ \delta {\bf v}\label{a1}\end{equation}
Then in order for this Bohm-like theory to agree with standard quantum
theory in the same sense that Bohmian theory does, one must require that
${\bf v}_{bl}$ also satisfy Eq.~\ref{e5}; that is, one must require
\begin{equation}
  {\bf \nabla}\cdot(|\Psi |^{2}\delta {\bf v})=0.\label{a2}\end{equation}
Deotto and Ghirardi \cite{DGh} have shown that it is possible to choose
${\bf v}_{bl}$ in such a way as to satisfy several requirements that
one may reasonably expect, in particular what they call ``genuine''
Galilean covariance. I will consider a simplified form of the theory
suggested in \cite{DGh}; I will take
\begin{equation}
   \delta {\bf v}=\lambda ({\bf \nabla}|\Psi |^{2})\times  {\bf v}_{b},
   \label{a3}\end{equation}
where $\lambda $ is a constant and  ${\bf v}_{b}$ is still given by
Eq.~\ref{e4};  ${\bf v}_{bl}$ is then given by Eq.~\ref{a1}.
This  ${\bf v}_{bl}$ will certainly not satisfy all of the requirements
imposed in \cite{DGh}; I will argue in the next section that this makes
this simple example of a  Bohm-like theory  implausible, but not
demonstrably incorrect.  For now, I will proceed to discuss the consequences
of the choice (\ref{a3}). This choice does at least satisfy Eq.~\ref{a2};
to see that, note that ${\bf v}_{b}={\bf \nabla}S$ and that
${\bf \nabla}\cdot (|\Psi |^{2}{\bf \nabla}|\Psi|^{2}\times {\bf \nabla}S)$
vanishes identically for any $|\Psi|^2$ and any $S$.

It is possible to discuss the distribution $P(t)$ that this Bohm-like
theory will imply for the example given by Eq.~\ref{e2} without having to find
the trajectories explicitly.  If it were the case that the x-component
of ${\bf v}_{bl}$ were positive everywhere on the plane $S_0$ for all
times $t\geq 0$, we could conclude that $P(t)=Q(t)$, just as we did
in Eq.~\ref{e10} for the standard Bohmian theory. As we shall see below, 
if this condition on the x-component of ${\bf v}_{bl}$ is {\em not}
satisfied, then this Bohm-like theory will necessarily imply a 
{\em different} distribution $P(t)$ than does standard Bohmian theory. From
Eqs.~\ref{e2}, \ref{e7}, \ref{e8}, and \ref{a3}, the x-component of
$\delta {\bf v}$ at $X=0$ is
\begin{equation}
  \delta v_{x}(0,Y,Z,t)=2\lambda|\Psi(0,Y,Z.t)|^{2}(c^{2}-b^{2})YZt
  /(|\beta|^{4}|\gamma|^{4}),\label{a4}\end{equation}
while from Eq.~\ref{e6},
\begin{equation}
  v_{bx}(0,Y,Z,t)=(k+x_{1}t)/|\alpha|^{4}.\label{a5}\end{equation}
Let me now take $\lambda >0$ and $(c^{2}-b^{2})<0$.  Then in the two
quadrants of the plane $S_0$ with the product $YZ$ negative, $\delta  v_{x}$
will be positive, and since $v_{bx}$ is positive everywhere on $S_0$,
we will have $v_{blx}\;(=v_{bx}+\delta v_{x})>0$. On the other hand, in the
quadrants with $YZ$ positive, $\delta v_{x}$ is negative, and so 
$v_{blx}$ will be positive if and only if
$|\delta v_{x}|\leq v_{bx}$.
For a fixed value of $t$, the maximum value of $|\delta v_{x}(0,X,Y,t)|$ occurs
at the points $Y=\pm |\beta|^{2}/(\sqrt{2}b)$, 
$Z=\pm |\gamma|^{2}/(\sqrt{2}c)$. This maximum value is
\begin{equation}
|\delta v_{x}|_{{\rm max}}=\lambda\frac{a(b^{2}-c^{2})t}{\pi^{\frac{3}{2}}
  e|\alpha|^{2}|\beta|^{4}|\gamma|^{4}}\exp\left[\frac{-a^{2}(x_{1}-kt)^{2}}
  {|\alpha|^{4}}\right].\label{a7}\end{equation}
From Eq.~\ref{a7}, $|\delta v_{x}|_{{\rm max}}$ is zero at $t=0$ and is
proportional to $t^{-4}$ as $t\rightarrow \infty$, while from Eq.~\ref{a5},
$v_{bx}$ is non-zero at $t=0$ and is proportional to $t^{-1}$ as
$t\rightarrow \infty$. Therefore it is possible to have a value of
$\lambda$ sufficiently small so that $|\delta v_{x}|_{{\rm max}}<v_{bx}$
for all times $t\geq0$. In that case, $v_{blx}$ would be positive
everywhere on $S_0$ for all $t\geq0$, and so the values of
$P(t)$ in this Bohm-like theory and in the standard Bohmian theory
would agree.

Now let me take $\lambda$ to be sufficiently large so that
{$|\delta v_{x}|_{{\rm max}}>v_{bx}$ for some time $t>0$.  
This means that, for some values of $Y$, $Z$, and $t$,
\mbox{$v_{blx}(0,Y,Z,t)<0$}, 
which implies that some members of the Bohmian
ensemble are returning from $S_+$ to $S_-$.
Let $t_r$ be within an interval of time in which this return
is occurring. At any time $t$, the fraction of the ensemble in $S_+$
equals $Q(t)$, but at $t_r$ there is an additional fraction of
``returned'' members, which are in $S_-$ at $t_r$ but were in $S_+$
at some time prior to $t_r$.  This means that $P(t_{r})$ (which is the
total fraction of ensemble members that were in $S_+$ 
at {\em any} time $t'\leq t$) must be greater than $Q(t_{r})$.

To be certain of this conclusion, we must show that, of the ensemble
members which returned from $S_+$ prior to $t_r$, at least
a finite fraction still are in $S_-$ at $t_r$.  Let $D$ be an open, bounded 
region of the  plane $S_0$, such that at every point of $D$ and for an
interval of time around $t_r$, $v_{blx}$ is negative; such a region
must exist, if $\lambda$ is sufficiently large.  For sufficiently small
$\epsilon $, it must be possible to find a subset $D_{\epsilon }\subset D$
such that the distance between any point in $D_{\epsilon}$ and any point
on $S_0$
not in $D$ is at least $\epsilon$. Now it can be shown that, for Y and Z
bounded, the magnitude of the component of ${\bf v}_{bl}$ parallel to
$S_0$ is bounded, independently of $X$ and $t$; call such a
bound $|v_{\parallel}|_{{\rm max}}$.  Thus any member of the ensemble 
which returns to  $S_-$ through $D_{\epsilon}$ must spent at least an
amount of time $\tau =\epsilon/|v_{\parallel}|_{{\rm max}}$ in $S_-$
(because it takes at least time $\tau$ for it to clear the region $D$).
Thus all members of the ensemble which return to $S_-$ through
$D_{\epsilon}$ in the time interval $[t_{r}-\tau ,\;  t_{r}]$
will still be in $S_-$ at time $t_r$.

We therefore see that, with the wave function as given in Eq.~\ref{e2},
the Bohm-like theory defined by Eq.~\ref{a3} with a sufficiently-large
value of $\lambda $ will imply that $P(t)>Q(t)$, for some values of $t$.
Since with this wave function the standard Bohmian theory gives 
$P(t)=Q(t)$ for all $t$, we conclude that these two theories can
give different arrival-time distributions $P(t)$.

\section{Discussion}   
The choice for $\delta {\bf v}$ made in Eq.~\ref{a3} does not respect many
of the conditions set out by Deotto and Ghirardi \cite{DGh}. For example,
the cross product of two vectors is a pseudo vector, although a velocity
must of course be a true vector.  To take this choice seriously, one would
have to say that Eq.~\ref{a3} is only valid in a particular coordinate
system; if you want to know  $\delta {\bf v}$ in some other coordinate
system, use Eq.~\ref{a3} to calculate it in the particular system, and
then transform.  Deotto and Ghirardi require that there not be any 
preferred coordinate system; while this requirement is certainly quite
reasonable, it is not, strictly speaking, necessary.  As long as the
Bohm-like theory reproduces the observational consequences of standard
quantum theory, the preferred coordinate system remains hidden; its
existence can be neither confirmed nor refuted by any experimental result.

Still, Deotto and Ghirardi, and in a different way Holland \cite{Holland},
have shown that is is possible to formulate a Bohm-like theory which is
considerably more plausible than the one defined by Eq.~\ref{a3}.
It is certainly an
important question for the program of studying Bohmian theories, to 
judge which of the possible alternatives for the particle velocity
is the most plausible.  However, plausibility is not an issue we must
be concerned with here. One could certainly criticize the calculations
presented here, because of the implausibility of the choice
(\ref{a3}) or for that matter because the condition $Q(0)=0$ is not strictly
satisfied.  The calculation presented here does have the virtue of 
simplicity, and it is hard to believe that the result obtained
(that $P(t)$ differs from that implied by the standard Bohmian theory)
is an artifact either of the transformation properties of Eq.~\ref{a3}
or of the (arbitrarily small) tail of the initial wave function.
Rather, this result gives one confidence to conjecture that for
{\em any} Bohm-like theory (with non-trivial  $\delta {\bf v}$)
there exists an example of a wave function with $Q(0)=0$ exactly,
for which that theory and standard Bohmian theory yield different $P(t)$.

The example presented here does not imply any additional ambiguity
within the Bohmian program, beyond that already recognized in \cite{DGh}
and \cite{Holland}.  It is obvious that, when theories make different
choices for $\delta {\bf v}$, there will be some quantities for which
those theories will imply different results. What this example does show
is that such theories will differ on a quantity, namely the distribution
of arrival times, that one might have hoped would be definable strictly in
terms of the wave function, and so would be independent of any particular
completion of standard quantum theory.

To discuss this matter further, let us consider to what extent the 
demonstration by
Leavens \cite{Leavens}, that results from Bohmian theory
disagree with those from the theory of Grot, Rovelli, and Tate \cite{GRT}, 
should be counted as evidence against the latter theory.   If
one is committed to believing in the truth of Bohmian theory, one will
consider as incorrect anything which disagrees with it.
But even if one has no such commitment, one might be at least suspicious
of any result expressed solely in terms of the wave function
which disagrees with the Bohmian result, not because the
Bohmian theory is {\em necessarily} correct, but just because it
{\em might} be. Certainly there are quantities which can be precisely
defined and calculated within Bohmian theory, to which standard quantum 
theory assigns no meaning.  However, to the extent that
Bohmian theory can be considered a completion of standard quantum theory,
one might expect that, for any quantity that {\em can} be calculated
within standard quantum theory, the Bohmian calculation would agree.
Similarly, one might expect any quantity calculable from the
wave function to agree with {\em all} completions (including any
that might be judged implausible). As we have seen, for the distribution
of arrival times this is not possible.

Of course, if one asks about the results of a particular experiment designed
to measure times of arrival, quantum theory should be able to     
give an unambiguous answer, and Bohmian theory as well as any Bohm-like
theory should agree with that answer.  The issue we are considering is 
whether that answer can be stated, within standard quantum theory, in a 
way which is independent of the particular way in which the arrival times
are to be measured.  In standard quantum theory, no result is meaningful
unless it is measured; the quantity $Q(t)$ defined in Eq.~\ref{ez} must be
interpreted as the probability that the particle be found in $S_+$
at time $t$, rather than the probability that it {\em is}
there.  Nevertheless, we do not have to consider the particular way in 
which the particle's position is measured; in terms of Bohm-like theories,
we can say that they all must agree on the quantity $Q(t)$.  One might
have thought that $P(t)$ would enjoy the same status; after all, $P(t)$ is,
roughly speaking, like the disjunction of $Q(t')$ for $0\leq t' \leq t$.
Unfortunately, a determination of position at one time will disturb
the determination at any other time, and different Bohm-like 
theories, while constrained to have identical ensembles of positions
at any one time, differ precisely because they have different trajectories.

The ambiguity in the time-of-arrival distribution 
revealed by the example discussed
here would not be present in a one-dimensional example.  The 
analogue of Eq.~\ref{a2} for one dimension, namely $\partial(|\Psi|^{2}
\delta v)/\partial x=0$, together with suitable boundary conditions,
would require $\delta v=0$. One would expect that general arguments for
or against the definability of the arrival-time distribution, such as 
those quoted at the beginning of this paper, would be equally cogent
in one and in three dimensions. It does seem, however, that there
is an ambiguity in three dimensions which is not apparent if one only
considers one-dimensional examples.  Perhaps one dimension is
misleadingly simple; one may be tempted by the isomorphism between
configuration space and temporal space to ignore the special role
played by time in non-relativistic quantum mechanics.

The discussion above has been within the context of Bohmian and
Bohm-like theories, but many of the same points can be put entirely
within the context of standard quantum theory, by considering the
quantum probability current (to be denoted ${\bf J}$).  
As Squires has pointed out \cite{Squires}, the freedom to chose
alternative expressions for the velocity in Bohm-like theories is
a direct reflection of the under-determination of the quantum probability
current in more than one spatial dimension. Mielnik \cite{Mielnik}
suggested that a reasonable first guess for a time-of-arrival
density would be the component of ${\bf J}$ normal to the arrival
surface; in our case this would mean identifying $J_{x}(x=0,y,z,t)$
as the probability for arriving at the point $(0,y,z)$ at time $t$.
Mielnik then went on to show that this could not be correct in general,
since there must exist examples in which this component becomes negative.
It is sometimes suggested (for example, in \cite{DM}) that $J_{x}$
does indeed give the correct arrival time density, in those cases in
which it is always positive.

Let ${\bf J}_c$ denote the customary form for the 
quantum probability current  (which is just the product
of $|\Psi|^2$ with ${\bf v}_b$ which is given in Eq.~\ref{e4}); then
without now identifying  ${\bf v}_b$ as the velocity of anything, we
can see from Eq.~\ref{e9} that for the wave function given in Eq.~\ref{e2},
$J_{cx}$ is indeed positive everywhere on the  plane $S_0$.  So, if we
follow the above suggestion, we would say that $J_{cx}(x=0,y,z,t)$
does indeed give the arrival-time density for this wave function.

Now define ${\bf J}_{l}$ to be $(|\Psi|^{2}{\bf v}_{bl})$, 
where ${\bf v}_{bl}$
is given by Eqs.~\ref{a1} and \ref{a3} (and also need not be identified
as the velocity of anything). Then from Eqs.~\ref{e5} and \ref{a2} it follows
that
\begin{equation}
  {\bf \nabla}\cdot {\bf J}_{l}=-\frac{\partial |\Psi|^{2}}
  {\partial t}, \end{equation}
which means that we can, if we wish, violate custom and call
${\bf J}_{l}$ (instead of ${\bf J}_{c}$) the quantum probability current.
So we might as well say that $J_{lx}$ gives the arrival-time density,
in those cases in which $J_{lx}$ is always positive.

The calculations of the previous section show that, for the wave function
given in Eq.~\ref{e2}, if $\lambda$ happens to be small enough, 
then $J_{lx}$ is positive everywhere on $S_0$. So, for small
enough $\lambda$, the two possibilities for the probability current 
(${\bf J}_c$ and ${\bf J}_{l}$) give us two possibilities for the
arrival-time density ($J_{cx}$ and $J_{lx})$ which disagree 
with each other \cite{int}.  One certainly can make an arbitrary
choice between  ${\bf J}_c$ and ${\bf J}_{l}$, that is, one can pick
either one of them and choose to define that one to be 
the probability current. That choice, however, does not have any experimental 
implications---no experimental result can depend upon which definition
one happens to make---so it would not make sense to expect the choice
to be either confirmed or refuted by any experiment.  More generally,
one certainly can (and sometimes one does \cite{GRT}) identify some
quantity which can be calculated purely in terms of the wave function,
and choose to call that quantity a time-of-arrival distribution. 
One can then discuss the question of whether this quantity does
have \cite{GRT} or does not have \cite{ORU} properties that one might
intuitively expect such a distribution to have.  However, one should not
then expect that any experiment will confirm the felicity of that choice.

\section*{Acknowledgement}
I  acknowledge the hospitality of the
Lawrence Berkeley National Laboratory.


\vspace{2cm}
\begin{thebibliography}{99}
\bibitem{aha}Y. Aharonov, J. Oppenheim, S. Popescu, B. Reznik, and W. G. 
        Unruh, Phys. Rev. A {\bf 57}, 4130 (1998).
\bibitem{Allcock}G. R. Allcock, Ann. Phys. (N.Y.) {\bf 53} 253 (1969);
        {\bf 53} 286 (1969); {\bf 53} 311 (1969).
\bibitem{Mielnik}B. Mielnik, Found. Phys. {\bf 24}, 1113 (1994).
\bibitem{HZ}J. J. Halliwell and E. Zafiris, Phys. Rev. D {\bf 57}, 3351 
        (1998).
\bibitem{Del1}V. Delgado, Phys. Rev. A {\bf 57}, 762 (1998).
\bibitem{MBM}J. G. Muga, S. Brouard, and D. Mac\'{\i}as, Ann. Phys.
        (N.Y.) {\bf 240}, 351 (1995).
\bibitem{GRT} N. Grot, C. Rovelli, and R. S. Tate, Phys. Rev. A {\bf 54},
        4676 (1996).
\bibitem{DM}V. Delgado and J. G. Muga, phys. Rev. A {\bf 56}, 3425 (1997).
\bibitem{Del2}V. Delgado, preprint quant-ph/9805058 (1998).
\bibitem{Bohm}D. Bohm, Phys. Rev. {\bf 85}, 166 (1952); {\bf 85}, 180 (1952).
\bibitem{BH}D. Bohm and B. Hiley, {\it The Undivided Universe: An Ontological
        Interpretation of Quantum Mechanics} (Routledge, London, 1993).
\bibitem{Leavens}C. R. Leavens, Phys. Rev. A {\bf 58}, 840 (1998).
\bibitem{DGh}E. Deotto and G. C. Ghirardi, Found. Phys. {\bf 28}, 1 (1998).
\bibitem{Holland}P. R. Holland, Found. Phys. {\bf 28}, 881 (1998).
\bibitem{ex}More accurately, one needs the generalization of Eqs.~\ref{e4}
        and \ref{e5} to systems of many particles.
\bibitem{Squires}E. J. Squires, in {\it Bohmian Mechanics and Quantum Theory:
        An Appraisal}, edited by J. T. Cushing, A. Fine, and S. Goldstein
        (Kluwer, Dordrecht, 1996), p. 131.
\bibitem{int}Of course these two possibilities do agree when integrated
        over the $yz$ plane; they disagree as densities.
\bibitem{ORU}J. Oppenheim, B. Reznik, and W. G. Unruh, ``Time-of Arrival
        States,'' preprint quant-ph/9807043 (1998).
\end{thebibliography}
\end{document}